\documentclass[sigplan,nonacm]{acmart}
\makeatletter
\def\@ACM@checkaffil{
  \if@ACM@instpresent\else
  \ClassWarningNoLine{\@classname}{No institution present for an affiliation}%
  \fi
  \if@ACM@citypresent\else
  \ClassWarningNoLine{\@classname}{No city present for an affiliation}%
  \fi
  \if@ACM@countrypresent\else
  \ClassWarningNoLine{\@classname}{No country present for an affiliation}%
  \fi
}
\makeatother

\usepackage{centernot}
\usepackage{bussproofs}
\usepackage{amsmath}
\usepackage{mathpartir}
\DeclareMathAlphabet{\mathcal}{OMS}{cmsy}{m}{n}
\SetMathAlphabet{\mathcal}{bold}{OMS}{cmsy}{b}{n}

\usepackage[os=mac]{menukeys}
\usepackage{pgfplots}
\usepackage{pgfplotstable}
\pgfplotsset{compat=1.18}
\usepgfplotslibrary{groupplots}

\usepackage{xcolor}
\usepackage{soul}
\usepackage{mathtools}

\definecolor{Light}{gray}{.90}
\sethlcolor{Light}

\usepackage[most]{tcolorbox}

\definecolor{oiBlue}{HTML}{0072B2}
\definecolor{oiGreen}{HTML}{009E73}
\definecolor{oiOrange}{HTML}{E69F00}
\definecolor{oiVermilion}{HTML}{D55E00}
\definecolor{oiBrown}{HTML}{8C564B}
\definecolor{oiPurple}{HTML}{CC79A7}
\definecolor{oiSky}{HTML}{56B4E9}
\definecolor{oiYellow}{HTML}{F0E442}
\definecolor{oiBlack}{HTML}{000000}

\usepackage{algorithm}
\usepackage[noend]{algpseudocode}
\usepackage{algorithmicx}

\usepackage[T1]{fontenc}
\tcbuselibrary{listings,breakable,skins}
\usepackage{listings}

\lstdefinelanguage{WordRASP}{
  morekeywords={ipt,scr,opt,W},
  morekeywords=[2]{fun,ife,whl,hlt},
  sensitive=true,
  morecomment=[l]{//},
  morestring=[b]",
}

\lstdefinestyle{wordraspstyle}{
  language=WordRASP,
  basicstyle=\ttfamily\footnotesize,
  keywordstyle=\color{blue!70!black}\bfseries,
  keywordstyle=[2]\color{orange!70!black}\bfseries,
  commentstyle=\color{green!40!black}\itshape,
  stringstyle=\color{orange!90!black},
  showstringspaces=false,
  showspaces=false,
  showtabs=false,
  breaklines=true,
  breakatwhitespace=false,
  columns=fullflexible,
  keepspaces=true,
  tabsize=2,
  upquote=true,
}

\newtcblisting{dslblock}[1][]{%
  breakable,
  enhanced,
  colback=black!1,
  colframe=black!15,
  boxrule=0.3pt,
  arc=2pt,
  left=6pt,
  right=2pt,
  top=4pt,
  bottom=4pt,
  listing only,
  listing options={style=wordraspstyle},
  title={#1},
  coltitle=black,
  fonttitle=\bfseries\small,
  boxed title style={
    colback=black!5,
    colframe=black!15,
    boxrule=0.3pt,
    arc=2pt
  },
  attach boxed title to top left={xshift=2mm,yshift*=-2mm},
}

\pgfplotsset{
  journal/.style={
    width=0.54\columnwidth,
    height=0.45\columnwidth,
    ymin=0,
    ylabel style={yshift=-3pt},
    enlarge x limits=0.10,
    axis line style={draw=black!45, line width=0.35pt},
    tick style={draw=black!45, line width=0.35pt},
    tick align=outside,
    ymajorgrids=true,
    xmajorgrids=false,
    grid style={draw=black!8, line width=0.25pt},
    tick label style={font=\scriptsize},
    label style={font=\scriptsize},
    title style={font=\small},
    every axis plot/.append style={draw=none},
  },
  journal legend/.style={
    legend style={
      font=\fontsize{5.2}{6}\selectfont,
      at={(0.02,0.98)},
      anchor=north west,
      draw=none,
      draw=none,
      fill=none,
      text opacity=1,
      row sep=0pt,
      inner xsep=1pt,
      inner ysep=1pt,
      nodes={scale=0.9, transform shape},
      legend columns=3
    },
    legend image post style={fill, draw=gray}
  }
}

\newtcolorbox{semanticsbox}{
  enhanced,
  colback=white,
  colframe=black,
  boxrule=0.5pt,
  arc=0pt,
  left=4pt,
  right=4pt,
  top=6pt,
  bottom=4pt,
  boxed title style={empty},
  coltitle=black,
}

\usepackage{wrapfig}

\usepackage{cuted}
\usepackage{stmaryrd}

\makeatletter
{%
  \begin{strip}
  \refstepcounter{widealgorithm}\label{#2}%
  \addcontentsline{loa}{widealgorithm}{\protect\numberline{\thealgorithm}{#1}}%
  \begingroup
  \fs@ruled 
  \@fs@pre
  \floatc@ruled{\fname@algorithm~\thealgorithm}{#1}\par
  \@fs@mid
  \begin{algorithmic}[1]
}
{%
  \end{algorithmic}
  \par\@fs@post
  \endgroup
  \end{strip}
}
\makeatother

\begin{document}

\author{\href{bre@ndan.co}{Breandan Considine}}

\title{Towards a Linear-Algebraic Hypervisor}

\begin{abstract}
  Many techniques in program synthesis, superoptimization, and array programming require parallel rollouts of general-purpose programs. GPUs, while capable targets for domain-specific parallelism, are traditionally underutilized by such workloads. Motivated by this opportunity, we introduce a pleasingly parallel virtual machine and benchmark its performance by evaluating millions of concurrent array programs, observing speedups up to $147\times$ relative to serial evaluation.
\end{abstract}

\maketitle

\section{Introduction}

An \textit{abstract machine} is a sequential model of computation for analyzing the complexity of algorithms, which, unlike the Turing machine, more closely characterizes the operational behavior of modern computers. Elgot and Robinson~\cite{elgot1964random} give an early example of such a machine, known as the random-access stored program (RASP) machine, equipped with a small set of atomic instructions and a shared address space storing both program and data. Cook and Reckhow~\cite{cook1972time} later present a simplified RASP and introduce an equivalent model which isolates its program from its data in memory, known as the random-access machine (RAM). An implementation of an abstract machine is known as a \textit{virtual machine} (VM) and a \textit{hypervisor} is a virtualization environment that simulates multiple VMs on a single physical machine, such as a GPU.

\subsection{RASP model}

A RASP machine is a 4-tuple \(\mathcal{R} = \langle C,c_0,\to,F\rangle\) consisting of a configuration space, \(C=\mathbb N\times\mathbb Z\times\mathbb Z^{\mathbb N}\times\mathbb Z^*\times\mathbb Z^*\), which we will write as $\langle i,a,M,u,y\rangle: C$, with \(i\) being the \textit{instruction counter}, \(a\) being the \textit{accumulator}, \(M\) being the \textit{memory}, \(u\) being the \textit{unread input}, and \(y\) being the \textit{output}. A RASP program is a vector $P: (\mathbb Z\times\mathbb Z)^m$ that is packed with an \textit{input} $x: \mathbb{Z}^*$, into an \textit{initial configuration}, \(c_0(P, x)\) as \(\langle 0,0,M'_{0..2m} \gets P,x0,\varepsilon \rangle\). Letting \(\langle o,j\rangle=\langle M_i,M_{i+1}\rangle\) denote the \textit{opcode} and \textit{operand} respectively, the one-step transition function $\rightarrow\::C^C$ is
\[\small
  \langle i,a,M,u,y\rangle \rightarrow
  \begin{cases}
    \langle i+2,\ j    ,\ M               ,\ u,\ y          \rangle, & o=1_\texttt{LOD},\\
    \langle i+2,\ a+M_j,\ M               ,\ u,\ y          \rangle, & o=2_\texttt{ADD},\\
    \langle i+2,\ a-M_j,\ M               ,\ u,\ y          \rangle, & o=3_\texttt{SUB},\\
    \langle i+2,\ a    ,\ M_j\gets a      ,\ u,\ y          \rangle, & o=4_\texttt{STO},\\
    \langle j  ,\ a    ,\ M               ,\ u,\ y          \rangle, & o=5_\texttt{BPA}\ \wedge\ a>0,\\
    \langle i+2,\ a    ,\ M               ,\ u,\ y          \rangle, & o=5_\texttt{BPA}\ \wedge\ a\le 0,\\
    \langle i+2,\ a    ,\ M_j\gets \sigma ,\ v,\ y          \rangle, & o=6_\texttt{RD}\ \wedge\ u=\sigma v,\\
    \langle i+2,\ a    ,\ M               ,\ u,\ y\cdot M_j \rangle, & o=7_\texttt{PRI},\\
    \langle i  ,\ a    ,\ M               ,\ u,\ y          \rangle, & \text{otherwise.}
  \end{cases}
\]

\noindent The RASP machine's halting configurations, $F \subseteq C$, are characterized by the fixed points, $F= \{c \in C \mid c \rightarrow c\}$.
\pagebreak

\section{Method}

We will consider a parallel RASP variant loosely equivalent to the PRAM~\cite{fortune1978parallelism} model, except each machine stores its program in memory and all machines occupy constant space.

Concretely, we lift the transition function to a vectorized transition map, \(\Phi:C^d\to C^d\) whose iterates are given by \( \Phi^{i+1}=\Phi\circ\Phi^i\), so the resulting configuration after \(i\) steps is \( \mathbf c_i=\Phi^i(\mathbf c_0).\) Thus, for a vector of programs \(\mathbf{P}\) and inputs \(\mathbf x\), we will write \( \mathbf c_i(\mathbf P,\mathbf x):=\Phi^i\bigl(\mathbf c_0(\mathbf P,\mathbf x)\bigr)\) for the evolution of $n$ independent machines, each with static memory. Since memory is statically allocated, each stepwise transition is a Boolean function on finitely many words, hence $\Phi$ admits a unique multilinear polynomial representation over \(\mathbb F_2\)~\cite{mossel2003learning}. 

\subsection{Word RASP}

Fix \(w,m,n,\ell,s\in\mathbb N\), and assume a fixed-length word, \(\mathbb W=\mathbb F_2^w \). Then, for a program \(P : \mathbb W^{2m\leq n}\), define the word RASP as \( \mathcal R^{(w,n,\ell,s)}=\langle C_w,c_0,\to,F\rangle \) with a finite configuration space,
\[
  \setlength{\arraycolsep}{2pt}
  \begin{array}{c c c c c c c c c c c c}
    \phantom{.......}C_w & = & \mathbb W & \times & \mathbb W & \times & \mathbb W^n & \times & \mathbb W^{\ell+1} & \times & \mathbb W^{s+1} & = \mathbb F_2^{\,w(n+\ell+s+4)},\\
    \phantom{.......}c   & = & \langle i & , & a & , & M & , & u & , & y \rangle &
  \end{array}\vspace{0.1cm}
\]
\noindent where the initial configuration on input $x$ is given by
\[
  c_0(P, x) =
  \big\langle
  0_w,\,
  0_w,\,
  P\,0_w^{\,n-2m},\,
  x0_w^{\ell-|x|},\,
  0_w^{s+1}
  \big\rangle. \phantom{extra extr space}\vspace{0.1cm}
\]

\noindent We write \( \oplus, \otimes: \mathbb{W} \times \mathbb{W} \rightarrow \mathbb{W}\) for addition and multiplication \(\pmod{2^w}\) and assume a circular memory addressing scheme,
\[
  \langle o,j\rangle:=\langle M_{\, i\bmod n},M_{\,( i\,\oplus 1)\bmod n}\rangle.\qquad
\]
Now, let $\delta_{jk}$ be the Kronecker delta and define the operators,
\[
  M_j \gets a := \big[\delta_{jk} a + (1-\delta_{jk})M_k\big]_{k=0}^{n-1},
\]
\[
  u^\bullet:=u_{\min(u_0 + 1,\ell)},
  \:\:\:
  u^\triangleright:=u_0 \gets \min( u_0 + 1,\ell),
\]
\[
  y :: z:=
  \begin{cases}
    (y_{y_0 + 1} \gets z)_{0} \gets y_0 + 1,
    &  y_0<s,\\
    y, & \text{otherwise.}
  \end{cases}
\]

\noindent The one-step transition operator takes $\langle i, a, M, u, y\rangle$ to
\[
  \small
  \begin{cases}
    \langle i\oplus 2_w,\ j,\ M,\ u,\ y\rangle,
    & o=1_\texttt{LOD},\\
    \langle i\oplus 2_w,\ a\oplus M_{j\bmod n},\ M,\ u,\ y\rangle,
    & o=2_\texttt{ADD},\\
    \langle i\oplus 2_w,\ a\otimes M_{j\bmod n},\ M,\ u,\ y\rangle,
    & o=3_\texttt{MUL},\\
    \langle i\oplus 2_w,\ a,\ M_{j\bmod n}\gets a,\ u,\ y\rangle,
    & o=4_\texttt{STO},\\
    \langle j,\ a,\ M,\ u,\ y\rangle,
    & o=5_\texttt{BNZ}\ \wedge\ a \neq 0,\\
    \langle i\oplus 2_w,\ a,\ M,\ u,\ y\rangle,
    & o=5_\texttt{BNZ}\ \wedge\ a = 0,\\
    \langle i\oplus 2_w,\ a,\ M_{j\bmod n}\gets u^\bullet,\ u^\triangleright,\ y\rangle,
    & o=6_\texttt{RD}\ \wedge\ u_0< \ell,\\
    \big\langle i\oplus 2_w,\ a,\ M,\ u,\ y :: M_{j\bmod n}\big\rangle,
    & o=7_\texttt{PRI},\\
    \langle i,\ a,\ M,\ u,\ y\rangle,
    & \text{otherwise.}\vspace{0.1cm}
  \end{cases}
\]
\noindent Once again, the word RASP machine's halting configurations are characterized by the fixed points, $F = \{c \in C_w \mid c \rightarrow c\}$.

Let \( \Phi_w :C_w\to C_w \) denote the resulting coordinatewise polynomial, so \(c\to c' \iff \Phi(c)=c'\). Alternatively, this can be written as a total deterministic transition function,
\[
\Phi_w(c)=\bigoplus_{\psi}\eta_\psi(c)\,\Phi_\psi(c), \qquad \bigoplus_{\psi}\eta_\psi(c) = 1,
\]
where \(\psi\) ranges over the guarded instruction cases, \(\eta_\psi\) is a multilinear indicator, and \(\Phi_\psi\) is the associated branch update.

\subsection{Heapless array lowering}\label{sec:heapless}

Consider a small array language with the following terms:
\[\small
  \begin{aligned}
    \texttt{F } &::= \texttt{ fun f0 ( ipt : W \char`\^\ N ) -> W \char`\^\ N \{ B \} } \\
    \texttt{N } &::= \texttt{ 1 } \mid \texttt{ 2 }  \mid \texttt{ 3 } \mid \texttt{ 4 } \mid \texttt{ 5 } \mid \texttt{ 6 } \mid \texttt{ 7 } \mid \texttt{ 8 } \mid \texttt{ 9 } \\
    \texttt{B } &::= \texttt{ S } \mid \texttt{ S ; B } \mid \texttt{ hlt} \\
    \texttt{S } &::= \texttt{ P = E } \mid \texttt{ ife G \{ B \} \{ B \} } \mid \texttt{ whl G \{ B \}}\\
    \texttt{Z } &::= \texttt{ 0 } \mid \texttt{ N}\\
    \texttt{G } &::= \texttt{ ipt [ Z ] } \mid \texttt{ scr [ Z ] } \mid \texttt{ opt [ Z ]} \\
    \texttt{P } &::= \texttt{ scr [ Z ] } \mid \texttt{ opt [ Z ]} \\
    \texttt{E } &::= \texttt{ G + G } \mid \texttt{ G * G } \mid \texttt{ G + Z } \mid \texttt{ G * Z } \mid \texttt{ G } \mid \texttt{ Z }\\
  \end{aligned}
\]

\noindent Reserve disjoint memory regions \(p_{0\dots l-1}\), \(q_{0\dots e-1}\), and \(r_{0\dots \gamma-1}\) for arrays $\texttt{ipt}: \mathbb{W}^{l \le \ell}, \texttt{opt}: \mathbb{W}^{e \le s}$ and $\texttt{scr}: \mathbb{W}^{\gamma \le \mu}$, with $(5+\ell+s+\mu+2m)\le n$ and interpret array indexing modulo declared length, so \(\llbracket\alpha\rrbracket_z := M_{2m + \llbracket\alpha\rrbracket + (z\bmod |\alpha|)}\). We will now assign a small-step operational semantics to this language:

\begin{semanticsbox}
  \vspace{-.3cm}
  \begin{scriptsize}
    \begin{mathpar}
      \inferrule*[right=\scriptsize\textsc{Fun}]
      {\Gamma \vdash B \Rightarrow C}
      {\vdash \texttt{fun f0 ( ipt : I \char`\^\ N ) -> I \char`\^\ }l\texttt{ \{ }B\texttt{ \}}
      \Rightarrow C\:\keys{\texttt{PRI }$q_i$\!}\,{}_{i=0}^{l-1}\:\keys{\texttt{HLT}}}

      \inferrule*[right=\scriptsize\textsc{Get}]
      { }
      {\Gamma \vdash \alpha[z] \Rightarrow_a \keys{\texttt{LOD }$0$}\:\keys{\texttt{ADD }$\llbracket\alpha\rrbracket_z$\!}}
\hspace{0.2cm}
      \inferrule*[right=\scriptsize\textsc{Put}]
      {\Gamma \vdash E \Rightarrow_a A}
      {\Gamma \vdash \alpha[z] \texttt{ = } E \Rightarrow A\:\keys{\texttt{STO }$\llbracket\alpha\rrbracket_z$\!}}

      \inferrule*[right=\scriptsize\(\oplus\)]
      {\Gamma \vdash I_1 \Rightarrow_a J_1 \\
      \Gamma \vdash I_2 \Rightarrow_a J_2}
      {\Gamma \vdash I_1 + I_2 \Rightarrow_a J_1\:\keys{\texttt{STO }$\nu$}\:J_2\:\keys{\texttt{ADD }$\nu$}}
      \hspace{0.1cm}
      \inferrule*[right=\scriptsize\(\otimes\)]
      {\Gamma \vdash I_1 \Rightarrow_a J_1 \\
      \Gamma \vdash I_2 \Rightarrow_a J_2}
      {\Gamma \vdash I_1 \times I_2 \Rightarrow_a J_1\:\keys{\texttt{STO }$\nu$}\:J_2\:\keys{\texttt{MUL }$\nu$}}

      \inferrule*[right=\scriptsize\textsc{Ife}]
      {\Gamma \vdash I \Rightarrow_a J \\
      \Gamma \vdash B_1 \Rightarrow C_1 \\
      \Gamma \vdash B_2 \Rightarrow C_2}
      {\Gamma \vdash \texttt{ife }I\ \texttt{\{}B_1\texttt{\}}\ \texttt{\{}B_2\texttt{\}}
      \Rightarrow
      J\:\keys{\texttt{BNZ }$\ell_t$\!}\:C_2\:\keys{\texttt{JMP }$\ell_e$\!}\:{}^{\ell_t:}\,C_1\:{}^{\ell_e:}}

      \inferrule*[right=\scriptsize\textsc{While}]
      {\Gamma \vdash I \Rightarrow_a J \\
      \Gamma \vdash B \Rightarrow C}
      {\Gamma \vdash \texttt{whl }I\ \texttt{\{}B\texttt{\}}
      \Rightarrow
      {}^{\ell_h:}\,J\:\keys{\texttt{BNZ }$\ell_b$\!}\:\keys{\texttt{JMP }$\ell_e$\!}\:{}^{\ell_b:}\,C\:\keys{\texttt{JMP }$\ell_h$\!}\:{}^{\ell_e:}}

      \inferrule*[right=\scriptsize\textsc{Seq}]
      {\Gamma \vdash B \Rightarrow C_B \\
      \Gamma \vdash S \Rightarrow C_S}
      {\Gamma \vdash B\texttt{ ; }S \Rightarrow
      C_B\:C_S}
      \hspace{0.3cm}
      \inferrule*[right=\scriptsize\textsc{Halt}]
      { }
      {\Gamma \vdash \texttt{hlt} \Rightarrow
      \keys{\texttt{PRI }$q_i$\!}\,{}_{i=0}^{l-1}\:\keys{\texttt{HLT}}}
    \end{mathpar}
  \end{scriptsize}
\end{semanticsbox}

\noindent We write \(\Gamma \vdash E \Rightarrow A\) when the instruction sequence \(A\) implements \(E\), and \(\Gamma \vdash E \Rightarrow_a A\) when $A$ implements $E$ and leaves value $a$ in the accumulator. Let $\nu$ be a single reserved memory address and finally, desugar \keys{\texttt{JMP }$\ell$} as \keys{\texttt{LOD }$1$}\:\keys{\texttt{BNZ }$\ell$}, with \(\keys{\texttt{HLT}}\) being any reserved invalid instruction pair whose execution would trigger a fixed-point halt (e.g., \keys{0 0}).

\subsection{Implementation}

VMs are implemented using a mainly branchless strategy to reduce warp divergence on the GPU, and host threads are dispatched to VMs using a simple round-robin scheduler. A short supplemental, including hypervisor pseudocode and platform-specific details, may be found in Appendix~\ref{sec:impl}.\pagebreak

\section{Evaluation}

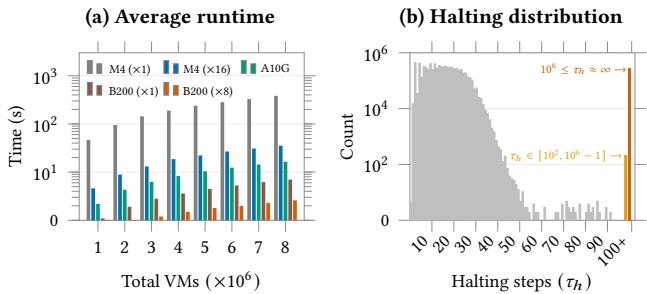
\begin{figure}[H]
  \hspace{-0.2cm}
  \begin{tikzpicture}
    \begin{groupplot}[group style={group size=2 by 1, horizontal sep=4em}]
      \nextgroupplot[
        journal,
        journal legend,
        title={\hspace{-0.3cm}\footnotesize\textbf{(a) Average runtime}},
        ylabel={Time (s)},
        ylabel style={yshift=-1pt},
        ymode=log,
        log basis y={10},
        ymin=1,
        ymax=3000,
        yticklabels={0,$10^1$,$10^2$,$10^3$,$10^4$},
        xtick={1,2,3,4,5,6,7,8},
        log ticks with fixed point,
        minor y tick num=4,
        yticklabel style={/pgf/number format/fixed},
        ybar=0.4pt,
        xlabel={Total VMs $(\times 10^6)$},
        bar width=1.4pt,
        legend image post style={scale=0.6}
      ]
      \addplot+[fill=gray] coordinates {
      (1,46.5)
      (2,95.2)
      (3,143.8)
      (4,188.9)
      (5,239.1)
      (6,282.0)
      (7,327.1)
      (8,382.6)
      };
      \addlegendentry{M4 ($\times1$)\phantom{a}}

      \addplot+[fill=oiBlue] coordinates {
        (1,4.6)
        (2,8.9)
        (3,13.1)
        (4,18.6)
        (5,22.2)
        (6,26.8)
        (7,30.8)
        (8,35.3)
      };
      \addlegendentry{M4 ($\times16$)\phantom{a}}

      \addplot+[fill=oiGreen] coordinates {
        (1,2.2)
        (2,4.3)
        (3,6.3)
        (4,8.3)
        (5,10.4)
        (6,12.4)
        (7,14.4)
        (8,16.5)
      };
      \addlegendentry{A10G}


      \addplot+[fill=oiBrown] coordinates {
        (1,1.1)
        (2,1.9)
        (3,2.8)
        (4,3.6)
        (5,4.5)
        (6,5.3)
        (7,6.2)
        (8,7.0)
      };
      \addlegendentry{\hspace{0.03cm}B200 ($\times1$)}

      \addplot+[fill=oiVermilion] coordinates {
        (1,0.6)
        (2,0.9)
        (3,1.2)
        (4,1.5)
        (5,1.8)
        (6,2.0)
        (7,2.3)
        (8,2.6)
      };
      \addlegendentry{\hspace{-0.05cm}B200 ($\times8$)}

\nextgroupplot[
journal,
title={\hspace{-0.3cm}\footnotesize\textbf{(b) Halting distribution}},
ylabel={Count},
xlabel={\vspace{-0.5cm}Halting steps $(\tau_{h})$},
xlabel shift=-6pt,
ybar,
ymode=log,
yticklabels={0,$10^2$,$10^4$,$10^6$},
log ticks with fixed point,
minor y tick num=4,
yticklabel style={/pgf/number format/fixed},
bar width=1pt,
ymin=1,
ymax=1000000,
xmin=0,
xmax=102.5,
enlarge x limits=false,
xtick={10,20,30,40,50,60,70,80,90,101},
xticklabels={10,20,30,40,50,60,70,80,90,100+},
xticklabel style={rotate=45, anchor=east},
]

      \addplot+[fill=lightgray] coordinates {
(1,1.2)
(2,2.2)
(3,3.1)
(4,4.1)
(5,5.1)
(6,6.0)
(7,7.0)
(8,8.0)
(1,0)
(2,0)
(3,0)
(4,0)
(5,15462)
(6,465523)
(7,36166)
(8,442713)
(9,139123)
(10,333902)
(11,305299)
(12,265647)
(13,421766)
(14,254552)
(15,417864)
(16,290138)
(17,364210)
(18,336307)
(19,320446)
(20,346245)
(21,301765)
(22,318766)
(23,300117)
(24,275325)
(25,292748)
(26,246539)
(27,264192)
(28,194893)
(29,195236)
(30,145839)
(31,131315)
(32,108096)
(33,51915)
(34,56625)
(35,26557)
(36,24251)
(37,14758)
(38,9242)
(39,7167)
(40,4155)
(41,2064)
(42,1528)
(43,753)
(44,445)
(45,353)
(46,129)
(47,207)
(48,75)
(49,40)
(50,32)
(51,28)
(52,20)
(53,13)
(54,5)
(55,11)
(56,3)
(57,3)
(58,2)
(59,1)
(60,4)
(61,2)
(62,2)
(63,0)
(64,1)
(65,3)
(66,0)
(67,1)
(68,1)
(69,1)
(70,2)
(71,2)
(72,0)
(73,4)
(74,1)
(75,5)
(76,3)
(77,2)
(78,3)
(79,2)
(80,0)
(81,4)
(82,0)
(83,1)
(84,2)
(85,4)
(86,2)
(87,3)
(88,1)
(89,5)
(90,2)
(91,1)
(92,0)
(93,3)
(94,0)
(95,2)
(96,0)
(97,1)
(98,1)
(99,1)
      };

\addplot+[fill=oiVermilion, draw=oiVermilion, bar width=0.7pt] coordinates { (100,269104) };
\addplot+[fill=oiOrange, draw=oiOrange, bar width=0.7pt] coordinates { (95,211) };
\node[oiOrange, align=right, anchor=east, font=\footnotesize] at (axis cs:101, 211) {\fontsize{4.6}{4}\selectfont$\tau_h\in [10^2,10^6-1]\to$};
\node[oiVermilion, align=right, anchor=east, font=\footnotesize] at (axis cs:103, 270000) {\fontsize{4.6}{4}\selectfont$10^6\le \tau_h \approx \infty\to$};

    \end{groupplot}
  \end{tikzpicture}
  \vspace{-0.6cm}
  \caption{Wallclock and simulation runtime of 8m programs.}
  \vspace{-0.3cm}
  \label{fig:rasp-eval}
\end{figure}

We implement VMs for the word RASP on (1)~Kotlin/JVM and (3) Nvidia CUDA with parameters $w=32$, $n~=~250$, $\ell~=~10$, $s=2$, $\tau_{\max}=10^6$. Then we evaluate each implementation on $d=8\times 10^6$ random array programs sampled uniformly from the space of valid code snippets of exactly length $L=100$ using Considine's word sampler~\cite{considine2025word}. In Fig.~\ref{fig:rasp-eval}a, we measure runtime across the JVM (running on an Apple M4 Max) and CUDA platforms (on the Nvidia A10G and B200 GPUs), reporting wall-clock time to evaluate up to $\tau_{\max}$ VM steps.
\vspace{-0.3cm}

\noindent In Fig.~\ref{fig:rasp-eval}b, we sample $d=8\times 10^6$ programs of length $L=100$ and estimate the halting probability before $\tau_{\max}=10^6$ steps, \[\Pr_{P' \sim \mathcal P_L}\Big(\Phi_w^{\tau_{\max}}\bigl(c_0(P'\Rightarrow P, x\sim \mathbb{F}_2^{w\cdot\text{arity}(P')}\bigr)\in F \bigm\vert L = |P'|\Big).\]
We plot the distribution of halting programs by their halting number $\tau_h$, i.e., the least $\tau$ such that $\Phi_w^\tau(c_0)=\Phi_w^{\tau+1}(c_0)$.

\section{Related work}

Prior work has explored compiling to transformers~\cite{weiss2021thinking}, though not intended as a practical compiler per se. The most relevant systems literature is Vectorvisor~\cite{ginzburg2023vectorvisor}, which targets WebAssembly IR, requiring a much larger set of opcodes. Exhaustive search has played a key role in recent empirical mathematics projects~\cite{blanchard2025determination, kauers2023flip} which use distributed computing to filter for constructions with favorable properties, but do not explicitly leverage SIMD architectures. We specifically target a much smaller IR and heap space, enabling us to run millions of concurrent VMs. The RASP model's simplicity, low memory footprint, and proximity to realistic assembly makes it suitable for algorithmic analysis, program synthesis, empirical research, and a variety of pedagogical applications.

\section{Conclusion}

We have presented a multilinear RASP VM, illustrated its viability by lowering a heapless array language, and used it to estimate the probability of sampling halting programs, demonstrating significant speedups over na\"ive evaluation. In future work, we intend to use it to accelerate the search for straight-line programs for fast matrix multiplication and context-free grammar parsing. Source code and data for all experiments will be released here: \href{https://github.com/breandan/lavm}{https://github.com/breandan/lavm}

\bibliography{../bib/acmart}
\appendix
\vspace{-0.2cm}
\section{Implementation details}\label{sec:impl}

We launch \(W\) persistent workers, where \(W\) is the maximum number of resident threads, and stripe the global configuration vector \(\mathbf c\), so that worker \(g\) visits each VM in its stripe in round-robin order. Each visit advances one live VM for a bounded epoch \(q\), yielding a static schedule that realizes the lifted dynamics up to a finite time horizon, \(\Phi^{\tau_{\max}}:C_w^d\to C_w^d\).
\vspace{-0.7cm}
\begin{algorithm}[H]
\caption{Striped round-robin RASP hypervisor}
\label{alg:hypervisor}
\small
\begin{algorithmic}[1]
\State \textbf{inputs:} Global config \(\mathbf c_0 \in C^d\), epoch \(q\), step budget \(\tau_{\max}\)
\State \(\mathbf c \gets \mathbf c_0\), \(W \gets \textsc{MaxResidentThreads}()\)
\For{\(g = 0\) \textbf{to} \(W-1\) \textbf{in parallel}} \Comment{Persistent workers}
\For{\(r = 0\) \textbf{to} \(\lceil \tau_{\max} / q \rceil - 1\)} \Comment{Total rounds}
\For{\(k = 0\) \textbf{to} \(\lceil d / W \rceil-1\)} \Comment{Walk worker stripe}
\State \(j \gets g + kW\) \Comment{Current VM index}
\If{\(j < d\) \textbf{and} \(\mathbf c[j] \notin F\)} \Comment{Skip halted VMs}
\State \(R \gets \textsc{LoadRegs}(\mathbf c[j])\)
\For{\(t = 1\) \textbf{to} \(q\)} \Comment{Bounded epoch}
\If{\(R \in F\)} \textbf{break} \EndIf
\State \(R \gets \Phi_w(R)\) \Comment{One-step transition}
\EndFor
\State \(\mathbf c[j] \gets \textsc{StoreRegs}(R)\)
\EndIf
\EndFor
\EndFor
\EndFor
\State \textbf{return} \(\mathbf c\)
\end{algorithmic}
\end{algorithm}

\vspace{-0.5cm}
\noindent Since CUDA is limited to at most 255 registers per thread~\cite{nvidia2026cuda}, we constrain VM memory to $n \lesssim 250$ 32-bit words, however this constraint can be relaxed, VRAM permitting, to 1 kB or higher, albeit at the cost of increased memory traffic. Assuming 256 resident threads per SM, an Nvidia B200 can sustain about 37,888 resident threads at peak theoretical occupancy.

\section{Heapless busy beavers $(L=120)$}

Listed below are a few of the busiest beavers discovered after conducting a $\sim\!20$ minute search with the JVM running on an M4 ($\times 16$) architecture using the grammar from Sec.~\ref{sec:heapless}, with the \texttt{ipt}, \texttt{opt} and \texttt{scr} parameters all initialized to 0s. \hspace{-0.3cm}

\begin{dslblock}[BB \#1 $(\tau_h = 1549)$]
   fun f0(ipt: W ^ 4) -> W ^ 1 {
      opt[0] = 4;
      whl opt[0] {
         opt[0] = opt[0] + opt[0];
         scr[3] = opt[0];
         whl opt[0] {
            whl opt[0] {
               scr[7] = scr[2] * 7;
               opt[0] = 0;
               scr[1] = 9
            }
         };
         opt[0] = scr[2] + 0;
         opt[0] = scr[3] + scr[5]
      }
   }
\end{dslblock}

\begin{dslblock}[BB \#2 $(\tau_h = 1272)$]
   fun f0(ipt: W ^ 2) -> W ^ 1 {
      opt[0] = opt[0] + 6;
      whl opt[0] {
         scr[3] = opt[0] + 0;
         ife ipt[0] { hlt } {
            scr[8] = 6;
            opt[0] = ipt[0] + 3; scr[0] = ipt[1] * 2;
            opt[0] = scr[3] * 6; scr[6] = scr[0] * 3;
            opt[0] = opt[0]
         }
      }
   }
\end{dslblock}

\begin{dslblock}[BB \#3 $(\tau_h = 1255)$]
   fun f0(ipt: W ^ 3) -> W ^ 1 {
      opt[0] = 1;
      whl opt[0] {
         opt[0] = opt[0] + ipt[0];
         ife ipt[2] { hlt } {
            scr[4] = 6;
            opt[0] = opt[0] + opt[0]
         };
         scr[6] = scr[8] * 2;
         whl ipt[1] { scr[4] = 3; opt[0] = 8 };
         scr[0] = opt[0] + 9
      }
   }
\end{dslblock}
\clearpage
\end{document}